\documentclass[letterpaper, 10pt, conference]{ieeeconf}  
\IEEEoverridecommandlockouts 
\usepackage{graphicx, subcaption, xcolor, cite} 
\usepackage{algorithm, algpseudocode}
\usepackage{booktabs, siunitx}
\usepackage{hyperref}
\hypersetup{colorlinks=true, breaklinks=true, plainpages=false, citecolor=blue, linkcolor=blue, urlcolor=NCSUorange}
\usepackage{amsmath, amssymb}
\definecolor{NCSUred}{RGB}{153, 0, 0}
\definecolor{NCSUgreen}{RGB}{0, 132, 115}
\definecolor{NCSUblue}{RGB}{65, 86, 161}
\definecolor{NCSUorange}{RGB}{209, 73, 5}

\newcommand{\mc}[1]{\mathcal{#1}}
\newcommand{\ms}[1]{\mathsf{#1}}
\newcommand{\mr}[1]{\mathrm{#1}}

\newcommand{\mbb}[1]{\mathbb{#1}}
\newcommand{\rg}[1]{\mathring{#1}}
\newcommand{\mR}{\mathbb{R}}

\newcommand{\mN}{\mathbb{N}}
\newcommand{\xD}[1]{\mr{d} #1}
\newcommand{\xDD}[2]{\frac{\xD{#1}}{\xD{#2}}}

\newcommand{\bra}[1]{\left( #1 \right)}

\newcommand{\BRA}[1]{\left\{ #1 \right\}}
\newcommand{\norm}[1]{\left\| #1 \right\|}
\newcommand{\ip}[2]{\langle #1, \, #2 \rangle}

\newcommand{\be}{\begin{equation}}
\newcommand{\ee}{\end{equation}}

\newtheorem{theorem}{Theorem}
\newtheorem{definition}{Definition}
\newtheorem{remark}{Remark}
\newtheorem{assumption}{Assumption}
\newtheorem{lemma}{Lemma}
\newtheorem{corollary}{Corollary}

\title{\LARGE \bf EDMD-Based Robust Observer Synthesis for Nonlinear Systems}
\author{Xiuzhen Ye, Wentao Tang
\thanks{This work was supported by NSF-CBET Award \#2414369 and ACS PRF Award \#66911-DNI9.}
\thanks{The authors are with Department of Chemical and Biomolecular Engineering, North Carolina State University, U.S.A. Corresponding author: W. Tang ({\tt\small wtang23@ncsu.edu})}
}

\begin{document}
\maketitle\thispagestyle{empty}
\pagestyle{empty}
\begin{abstract}
This paper presents a data-driven approach for designing state observers for continuous-time nonlinear systems, where an extended dynamic mode decomposition (EDMD) procedure is used to identify an approximate linear lifted model. 
Since such a model on a finite-dimensional space spanned by the dictionary functions has an inevitable mismatch, we first establish, based on our theory of reproducing kernel Hilbert space with a linear--radial kernel, that the nonlinear error magnitude in the approximate linear model is \emph{sectorially} bounded by the lifted state. 
The sector bound comprises a deterministic part due to the finite dictionary and a stochastic part due to the random data samples, and the observer design needs to account for both of these errors in a robust formulation. 
Hence, the observer synthesis is performed using linear matrix inequalities (LMIs), specified by the desired exponential decay rate of the observation error (when the system is asymptotically stable) or the $L^2$-gain from the modeling error to the observation error. 
Numerical studies demonstrate the effectiveness and flexibility of the proposed method. 
As such, this work entails an explicit elementary use of linear systems theory for nonlinear state observation in a Koopman operator-theoretic framework. 
\end{abstract}

%%%%%%%%%%%%%%%%%%%%%%%%%%%%%%%%%%%%%%%%%%%%%%%%%%%%%%%%%%%%%%%%%%%%%%%%%%%%%%%%
\section{Introduction}
\par Nonlinearity in the governing dynamics of engineering systems requires effective methods from nonlinear control theory \cite{TW_ACC_22, martin2023guarantees, de2023learning}, which in its latest developments features \emph{data-driven} modeling methods for analyzing and controlling nonlinear dynamics, such as those based on neural networks \cite{YR_CCE_22}, reinforcement learning \cite{Nian_Rev_CCE_20}, and Koopman modeling \cite{BP_ARC_21}. 
Typically, these methods assume full state information, e.g., in polynomial approximation-based controller design~\cite{MT_TAC_23}, kernel-based methods~\cite{FC_CDC_21}, and operator learning \cite{BS_arXiv_21}. 
Realistically, it is more likely that data access is limited to a (small) collection of measurable outputs. Hence, a {\it state observer} that infers state variables from measured variables is essential for the modeling, monitoring, and potential development of output-feedback control strategies. 
In a data-driven control context, naturally, the state observer in need should also be designed in a \emph{data-driven} manner. 

\par {\it Koopman operator} provides a powerful data-driven framework for the modeling, analysis, and control of nonlinear systems \cite{MA_Springer_20}. 
% From a modeling perspective, it captures nonlinear dynamics through a linear infinite-dimensional lifting \cite{}. 
Specifically, the Koopman operator (semigroup) maps state-dependent functions to their composition with the nonlinear flow, enabling nonlinear behaviors to be captured through linear evolution of the observables \cite{BS_arXiv_21}. 
Hence, by data-driven approximation of the Koopman operator, linear systems tools can potentially be applied to nonlinear systems -- such an idea has been widely applied to data-driven control problems in a practical manner \cite{son2021application, JK_CCE25, XYY_TCS_25}, while rigorous theoretical frameworks appeared more recently \cite{bevanda2024kernel, houska2025convex}. 

%IV need to discuss Ni and Tang
In this work, we focus on {\it Koopman-based observer design}. 
The idea of leveraging Koopman operators for observer synthesis was first introduced in~\cite{SB_IFAC_16}, where system states were lifted via Koopman eigenfunctions into a linear representation that admits a classical Luenberger observer. 
Using estimated Koopman eigenfunctions,~\cite{NT_arxiv_25} proposed a data-driven Kazantzis–Kravaris/Luenberger (KKL) observer for limit-cycle systems (for which the Koopman spectrum is discrete, with eigenfunctions spanning a rich function space). 
Similarly, a learning-based framework for observer synthesis in measure-preserving systems (including chaos) using Koopman operator theory was developed in~\cite{T_arxiv_25}. 
A general operator-theoretic perspective was provided in~\cite{MM_Arxiv_25}, which formulated Koopman-based Luenberger observer synthesis through operator equations and investigated its structural properties for systems on polydisks ($\mathbb{C}^n$). 

\par However, when the Koopman operator is approximated from data, especially when identified as a finite-rank one (i.e., a matrix) using a finite dictionary basis, \emph{the approximation error in the Koopman operator is inevitable}, and should be actively accounted for with a robust formulation. 
The characterization of such a \emph{mismatch}, however, does not seem to have been well addressed in the literature. 
Indeed, if the Koopman operator is well-defined in an infinite-dimensional space (such as $L^2$ \cite{NF_JNS_23} or a Sobolev space $H^s$ \cite{kohne2025error}), there would be no \emph{structural} mismatch, and the statistical error can be bounded by spectral analysis \cite{kostic2023sharp}. 
Yet, if a finite dictionary (typically, the dictionary of monomials up to a certain degree) is used,\footnote{
It remains a practical pathway to pursue Koopman modeling through finite dictionaries, either due to the lack of systematic theory for \emph{directly posing Koopman-based synthesis problems} in infinite-dimensional spaces, or for the interpretability of basis functions. In this paper, we consider the finite-dimensional ``elementary'' formulation and leave the ``advanced'' approach for future work.} 
then the structural mismatch needs to be captured by the projection from infinite dimensions to finite dimensions. 
A quantification of this projection error was given by Haseli and Cort{\'{e}}s \cite{haseli2023generalizing}, namely the invariance proximity, defined as the maximum relative residual over all test functions in the span of dictionary functions. This metric, however, tends to be conservative, since the worst test functions may not be fully relevant to the Koopman operator's capability to predict the state evolution. 

Hence, in this work, we establish an EDMD-based robust observer synthesis approach in the following way. 
\begin{itemize}
    \item We first establish a bound on the prediction error of the Koopman model of continuous-time autonomous systems, where the generator of Koopman semigroup is identified from data through generator extended dynamic mode decomposition (gEDMD) \cite{klus2020data}. 
    Using the reproducing kernel Hilbert space (RKHS) theory with a linear--radial kernel, introduced in the authors' earlier work \cite{TY_arxiv_25}, it is proved that the deterministic error takes a \emph{sectorial} form, which, after adding the sample-based stochastic error, remains sectorial. 
    \item Based on the ``gEDMD with sectorial error'' model, the robust observer synthesis problem is formulated as a semidefinite program (SDP). 
    The linear matrix inequality (LMI) constraint ensures that the state observation error decays exponentially at a prescribed rate despite the uncertainty, when the system is asymptotically stable, or when there is a bounded $L^2$-gain from the mismatch (nonlinear remainder) to the observation error. 
    The formulation resembles that of Str\"{a}sser et al. \cite{SR_TAC_24} for robust controller design under EDMD modeling error.\footnote{
    In \cite{SR_TAC_24}, however, it was assumed restrictively that the finite dictionary is closed and thus the sectorial error is only stochastic. Moreover, due to the bilinearity of the Koopman representation of control systems, the designed controller is (unnaturally) confined to a neighborhood of the origin. The present paper eliminates the closedness assumption on the dictionary. 
    }
\end{itemize}
The proposed observer is therefore applicable to nonlinear systems whenever the finite dictionary is rich enough to well approximate the infinite-dimensional lifting and the amount of data is sufficient to guarantee a low stochastic error. 
We evaluate the proposed observer design on three numerical examples, demonstrating the observer performance on systems with different state-space geometries. 

The rest of this paper is organized as follows. In \S\ref{sec_DataDrivenDynamicModelling}, we introduce the preliminaries of Koopman operator in dynamical systems, with error bound analysis on data-driven approximation. \S\ref{sec_KoopmanBasedObserver} provides the derivation of the proposed Koopman-based observer design with the resulting SDP. The numerical studies are presented in \S\ref{sec_numerical} followed by conclusions in \S\ref{sec_conclusion}.

\section{Data-Driven Modeling by gEDMD and Its Error Analysis}
\label{sec_DataDrivenDynamicModelling}
Consider a continuous-time nonlinear system:
\be\label{nonlinear_system}
\dot{x}(t) = f(x(t)), \quad y(t) = h(x(t)).
\ee
where $x(t) \in \mc{X}\subset \mathbb{R}^n$ and $y(t) \in \mathbb{R}^m$ denote the states and the outputs at time $t \geq 0$, respectively. 
\begin{assumption}
    Below we always assume that $\mc{X}$ is a compact region. The mapping $f: \mc{X} \rightarrow \mathbb{R}^n$ determines a semiflow on $\mc{X}$ over nonnegative times, which we denote by $x(t)=\mc{S}^t(x(0))$. 
    Without loss of generality, let the origin be an equilibrium point of \eqref{nonlinear_system}, i.e.,  $f(0) = 0$. 
\end{assumption}

\par The goal is to design an observer, such that the convergence of the error between the state and its estimate $e_x(t) : = x(t) - \hat{x}(t)$ is with an exponential rate, in which $\hat{x}(t)$ is the observed state. 
We suppose that the system dynamics $f$ is unknown, whereas a data sample from the system is available to allow the estimation of a surrogate model and the observer design based on such a surrogate.

\subsection{Koopman Operator Defined on a RKHS} 
Let us first review some preliminaries of Koopman operator theory. Under our setting for \eqref{nonlinear_system}, the Koopman semigroup $\{{\cal K}^t\}_{t\geq 0}$ is the semigroup specified by: 
\be\label{eq_def_Kt}
{\cal K}^t g: = g\circ {\cal S}^t, \enspace \forall \phi\in \mc{G}. 
\ee
That is, $\mc{K}^t g(x) = g(\mc{S}^t(x))$ for all state-dependent test functions $g$ in the (Banach) function space $\mc{G}$ of interest. 
For the definition to make sense, it is needed that $g\circ \mc{S}^t \in \mc{G}$ remains in $\mc{G}$, and we further require that the Koopman semigroup is strongly continuous, so that it possesses a closed and densely defined infinitesimal generator $\cal L$:
\be\label{eq_def_L}
{\cal L}g(x) : = \lim_{t \downarrow 0} \frac{1}{t}\left( {\cal K}^t g(x) - g(x) \right), \ \forall \phi \in D(\cal L),
\ee
where the domain of $\cal L$, $D(\cal L)$, is a closed subspace of $\mc{G}$ such that the above limit exists. Clearly, then, for any $g\in D(\mc{L}) \cap C^1(\mc X)$, ${\mc L}g = \nabla g\cdot f$.

\par Generally, the space $\mc{G}$ should be infinite-dimensional. But as we will use an EDMD-type approximation in a finite-dimensional subspace, it will have a deterministic error, and such an error must be accounted for in the observer synthesis. 
To facilitate the error analysis, it is desirable to choose the infinite-dimensional space $\mc{G}$ as a separable Hilbert space that is \emph{compactly embedded into $L^2(\mc X)$}, so as to hope for a ``minimal error'' when truncated to finite dimensions.
K{\"{o}}hne \cite{kohne2025error} considered the definition of (discrete-time) Koopman operator on Sobolev--Hilbert spaces $H^s(\mc X)$, which can be easily extended to a continuous-time (semigroup) setting. For such a definition, it is sufficient to require $f\in C^s(\mc X, \mR^n)$. 
However, as we argued in \cite{TY_arxiv_25}, the construction of $H^s(\mc X)$ does not entail the information of equilibrium point at the origin, and hence not yet a ``minimum'' space to accommodate the Koopman operator. 
If one uses a monomial dictionary in the form of $\{x_1, \cdots, x_n, x_1^2, x_1x_2, \cdots, x_n^2, \cdots \}$, then the dictionary does not ``cover'' the $H^s$ space in the high-degree limit. Hence, as in \cite{TY_arxiv_25}, we adopt an ``equilibrium-preserving'' RKHS formulation. 

\begin{definition}
    For $s\in \mbb{N}$, let $e_k: \mc X \to \mR$, $x\mapsto x_k$ ($k=1,\cdots, n$) be the component mappings. Denote
    $$\textstyle \rg H^s(\mc X) = \BRA{ \sum_{k=1}^n e_kg_k: g_k\in H^s(\mc X)}, \text{ and} $$
    $$\textstyle \rg C^s(\mc X, \mR^n) = \BRA{ \sum_{k=1}^n e_kg_k: g_k\in C^s(\mc X, \mR^n)}, $$
    as the space of ``locally at least linear'' $H^s$ (and $C^s$, respectively) functions. 
\end{definition}
\begin{lemma}
    If $f\in \rg{C}^s(\mc{X}, \mR^n)$ for some $s\in\mN$, then $\{\mc{K}^t\}_{t\geq 0}$ is a strongly continuous semigroup on $\rg{H}^s(\mc{X})$.     
\end{lemma} 
\begin{proof}
    For any $g\in H^s(\mc{X})$, the generalized derivative $\partial^\alpha (\mc{K}^tg)(x) = \partial^\alpha (g\circ \mc{S}^t)(x)$ with respect to any multi-index $\alpha$ of length $r\leq s$, by the chain rule, contains terms in the form of a product of $\partial^\beta g(\mc{S}^t(x))$ and partial derivatives of $\mc{S}^t$, where $\beta\leq \alpha$. In particular, the term with $\beta=\alpha$ is followed by $\partial_{\alpha_1} \mc{S}^t(x)\cdots \partial_{\alpha_r} \mc{S}^t(x)$, while every other term has a factor of higher-order derivatives of $\mc{S}^t$. 
    Under the given conditions, $\mc{S}^t \in C^s(\mc{X}, \mc{X})$ and at small $t$, $\mc{S} ^t(x)=x+O(t)$. 
    Therefore, $\partial^\alpha (\mc{K}^tg)\rightarrow \partial^\alpha g$ pointwise, and hence in $L^2$ due to the Lebesgue dominated convergence theorem, as $t\downarrow 0$. That is, $\lim_{t\downarrow 0} \mc{K}^tg = g$ in $H^s(\mc{X})$. 
    \par For $g\in \rg H^s(\mc{X})$, $g = \sum_{k=1}^n e_kg_k$ with $g_k\in H^s(\mc{X})$. Hence $\mc{K}^tg(x) = \sum_{k=1}^n \mc{S}^t_k(x) \mc{K}^tg_k(x)$, where $\mc{K}^tg_k \rightarrow g_k$ in $H^s(\mc{X})$ and $\mc{S}^t_k \rightarrow e_k$ in $C^s(\mc{X})$ as $t\downarrow 0$. Hence $\lim_{t\downarrow 0} \mc{K}^tg = g$ in $\rg H^s(\mc{X})$. 
\end{proof}

By defining an inner product on $\rg H^s(\mc X)$ as $\ip{e_jg}{e_kg'} = \delta_{jk}\ip{g}{g'}_{H^s}$, we see that $\rg H^s(\mc{X})$ is a Hilbert space. 
Moreover, since $H^s(\mc X)$ is a reproducing kernel Hilbert space (RKHS) whenever $s>n/2$, $\rg H^s(\mc X)$ is also an RKHS.\footnote{
    An RKHS is a Hilbert space in which the point evaluations $\delta_x: g\mapsto g(x)$ ($x\in \mc{X}$) are all continuous linear functionals and hence by Riesz-Fr{\'{e}}chet theorem, a member of the Hilbert space itself. For such a space, there always exists a so-called reproducing kernel function $\kappa: \mc{X}\times \mc{X}\to \mR$, which is symmetric and continuous, such that for any finite number of points $\{x_i\}_{i=1}^d \subset \mc{X}$, the matrix formed by $\kappa(x_i, x_j)$ is positive definite. On the RKHS, any $g$ satisfies $\ip{g}{\kappa(x,\cdot)} = g(x)$. We hence denote an RKHS with kernel $\kappa$ as $\mc{H}_\kappa$. 
}
\begin{lemma}[Tang and Ye \cite{TY_arxiv_25}]
\label{lem:linear--radial}
    The space $\rg H^s(\mc{X})$, when $s>n/2$, is equivalent to the RKHS with kernel $\rg\kappa (x, x') = (x\cdot x')\rho(\|x-x'\|)$, where $\rho: [0, \infty) \to [0, \infty)$ is such that its Fourier transform $\hat{\rho}$ satisfies $c_1(1+\|\xi\|^2)^{-s}\leq |\hat{\rho}(\xi)| \leq c_2(1+\|\xi\|^2)^{-s}$ for some constants $c_2\geq c_1>0$. We refer to $\rho$ as a radial function, and hence $\rg\kappa$ as a linear--radial kernel. 
\end{lemma}

\par On the RKHS, the kernel function with one argument fixed at $x\in \mc{X}$, $\rg\kappa(x,\cdot)=\rg\kappa_x \in \mc{H}_{\rg\kappa}$ is the lifted representation (i.e., canonical feature) of $x$. 
We note the following relation: $\ip{\rg\kappa_x}{\rg\kappa_{x'}} = \rg\kappa(x,x') = (x\cdot x')\rho(\|x-x'\|)$ and thus $\|\rg\kappa_x\| = \|x\| \rho(0)^{1/2} \propto \|x\|$. 
The action of the \emph{adjoint operator} of the Koopman operator $\mc{K}^t$ satisfies the following relation:
\begin{equation}\label{eq:push-forward}
    \textstyle \mc{K}^{t*} \rg\kappa(x, \cdot) = \rg\kappa(\mc{S}^t(x), \, \cdot) 
\end{equation}
and hence $\{\mc{K}^{t*}\}$ also forms a semigroup, i.e., the adjoint Koopman semigroup. Its infinitesimal generator is denoted as $\mc{L}^*$. The estimation of $\mc{L}$ or $\mc{L}^\ast$ therefore can be considered as a regression problem with respect to the equality \eqref{eq:push-forward}. 

\subsection{Deterministic Error of gEDMD at Large-Data Limit} 
\par EDMD is a common practice for data-driven approximation of the Koopman operator \cite{WK_JofNS_20, KN_PNP_20}, where a finite number of dictionary functions are used to restrict the operator onto their span. Therein, the EDMD error consists of both a projection error caused by the projection to this finite-dimensional subspace and a stochastic error due to the randomly drawn finite data sample \cite{NF_JNS_23, data_bound, KM_JNS_18}. 
Formally, when choosing a dictionary of $N$ linearly independent elements in $\mc{G} = \rg H^s(\mc{X}) = \mc{H}_{\rg\kappa}(\mc{X})$, an $N$-dimensional subspace $\mc{G}^N\subset \mc{G}$ is specified, and an approximate operator that plays the role of $\mc{L}^\ast$, say $\mc{L}_N^\ast$, is sought on $\mc{G}^N$. 
A common choice is $\mc{G}^N = \mr{span}\{\phi_k\}_{k=1}^N$, where $\phi_1 = e_1, \cdots, \phi_n = e_n, \phi_{n+1} = e_1^2, \phi_{n+2} = e_1e_2, \cdots$, until monomials up to a given degree are enumerated. This is indeed a consistent choice with $\mc{G} = \rg H^s(\mc{X})$, since the space of polynomials is dense in $\rg H^s(\mc{X})$. 
In this sense, the Koopman semigroup $\mc{K}^t$ defined on $\rg H^s(\mc{X})$ is comprehended as the limit of EDMD using an infinitely large dictionary \cite{WK_JofNS_20}.

% We require that $\phi_k(0) = 0$ and $\phi_k \in C^2({\cal X}, \mathbb{R})$. 
\par Denote the projection operator $\Pi_N: \mc{G}\to \mc{G}_N$ and the projected canonical feature of any $x\in \mc{X}$ as $\rg\kappa^N_x = \Pi_N\rg\kappa_x \in \mc{G}^N$.\footnote{Although $\rg\kappa^N$ may not be computable by elementary algebraic manipulations, there is no doubt about its conceptual existence and is actually the object such that for any $g\in \mc{G}^N$ and $x\in \mc{X}$, it holds that $\ip{g}{\rg\kappa^N_x}=g(x)$.} 
To obtain the best approximation on $\mc{G}^N$ from data, suppose that we have a sample of infinite size from a uniform probability distribution on $\mc{X}$: $\{(x, x^+=S^t(x))\}_{x\in \mc{X}}$. Then, by taking the optimum from:
\begin{equation}
\textstyle
    \min_{\mc{K}_N^{t*}: \mc{G}^N \to \mc{G}^N} \int_\mc{X} \| \mc{K}_N^{t*} \rg\kappa^N_{x} - \rg\kappa^N_{x^+} \|^2 \xD{x}, 
\end{equation}
we obtain the estimate for the $t$-interval adjoint Koopman operator at the large-data limit:
\begin{equation}\label{eq:EDMD.concept.dt}
    \textstyle 
    \mc{K}_N^{t*} = \bra{\int_{\mc{X}} \rg\kappa^N_{x^+}\times \rg\kappa^N_x\xD{x}} \bra{ \int_{\mc{X}} \rg\kappa^N_x\times \rg\kappa^N_x\xD{x} }^{-1}. 
\end{equation}
Here, on the Hilbert space $\mc{G}$, we use $g_1\times g_2$ to denote a rank-$1$ operator that maps any $g_3\in \mc{G}$ to $\ip{g_2}{g_3} g_1$.\footnote{
Because $\rg\kappa^N_{x^+} = \Pi_N \rg\kappa_{x^+} = \Pi_N \mc{K}^{t*} \rg\kappa_x$, we see that when $t\downarrow 0$, the result is formally reduced to: $\mc{L}_N^* = \Pi_N\mc{L}^*\bra{\int_{\mc{X}} \rg\kappa_x \times \rg\kappa^N_x\xD{x}} \bra{ \int_{\mc{X}} \rg\kappa^N_x\times \rg\kappa^N_x\xD{x} }^{-1}$, if we \emph{can} assume that $\forall x\in \mc{X}$, $\rg\kappa_x \in \mc{D}(\mc{L}^\ast)$ (which however may not be guaranteed). But the limit $t\downarrow 0$ can still be taken on the ``primal'' space, if the dictionary functions are $C^1$, as we see in the next subsection.
}

\par Now we analyze the deterministic error of $\mc{K}_N^{t\ast}$ \eqref{eq:EDMD.concept.dt} when used to \emph{predict the evolution of the projected canonical feature of any $x\in \mc{X}$}. That is, we need a bound on $\mc{K}_N^{t\ast} \rg\kappa^N_x - \rg\kappa^N_{x^+}$. 
It turns out that such an error bound is \emph{sectorial}, i.e., proportional to $\|\rg\kappa_x\|$ with a uniform ratio. 
\begin{theorem}[Deterministic error of prediction]\label{th:deterministic.error}
    Suppose that for all $N\in \mbb{N}$, there exists a $R_N>0$ independent of $x\in \mc{X}$, such that $\|(I-\Pi_N^*\Pi_N)\rg\kappa_x\|\leq R_N\|\rg\kappa_x\|$ ($\forall x\in \mc{X}$), and $C_N>0$ such that 
    $$\textstyle \norm{\bra{\int_{\mc{X}} \rg\kappa^N_x\times \rg\kappa^N_x\xD{x} }^{-1}} \leq C_N/ \int_\mc{X} \|\rg{\kappa}_x\|^2 \xD{x}.$$
    Then, given any $\psi\in D(\mc{L})\cap \mc{G}^N$, for all $t\geq 0$, $N\in \mbb{N}$, and $x\in \mc{X}$, we have 
    $$|\psi(\mc{S}^t(x))-\psi(x)| \leq \beta(\alpha^t-1)R_N(1+C_N)\|\rg\kappa_x\|\|\mc{L}\psi\|. $$
    for some $\alpha, \beta>0$. 
\end{theorem}
\begin{proof}
    Denote $\tilde{\mc{K}} = \mc{K}^t-I$ and $\tilde{\mc{K}}_N = \mc{K}_N^t-I$. We can verify that $\psi(\mc{S}^t(x))-\psi(x) = \ip{\mc{K}_N^{t\ast} \rg\kappa^N_x - \rg\kappa^N_{x^+}}{\psi} = \ip{\tilde{\mc{K}}_N^{t\ast} \Pi_N\rg\kappa_x - \Pi_N\tilde{\mc{K}}^{t\ast}\Pi_N^\ast \Pi_N\rg\kappa_x}{\psi} + \ip{\Pi_N \tilde{\mc{K}}^{t\ast} \Pi_N^\ast \Pi_N\rg\kappa_x - \Pi_N \tilde{\mc{K}}^{t\ast}\rg\kappa_x}{\psi} =: \mr{I} + \mr{II}$. Seeking bounds on the two terms:
$$\begin{footnotesize}
    \begin{aligned}
    |\mr{I}| & \overset{\eqref{eq:EDMD.concept.dt}}{=} 
    |\langle \Pi_N \tilde{\mc{K}}^{t\ast} \int_{\mc X} (I-\Pi_N^\ast\Pi_N)\rg\kappa_x \times \rg\kappa^N_x \xD{x} \bra{\int_{\mc{X}} \rg\kappa^N_x \times \rg\kappa^N_x \xD{x}} ^{-1} \Pi_N\rg\kappa_x , \, \psi\rangle| \\
    &\leq |\ip{\int_{\mc X} (I-\Pi_N^\ast\Pi_N)\rg\kappa_x \times \rg\kappa^N_x \xD{x} \bra{\int_{\mc{X}} \rg\kappa^N_x \times \rg\kappa^N_x \xD{x}}^{-1} \Pi_N\rg\kappa_x}{ \tilde{\mc{K}}^t \Pi_N^\ast \psi}| \\
    &\leq R_NC_N\|\Pi_N\rg\kappa_x\| \|\tilde{\mc{K}}^t\Pi_N^\ast\psi\|. 
\end{aligned}
\end{footnotesize}$$
$$\begin{footnotesize}
    \begin{aligned}
    |\mr{II}| &= |\ip{\Pi_N \tilde{\mc{K}}^{t*}(I-\Pi_N^\ast \Pi_N) \rg\kappa_x}{\psi}| = |\ip{(I-\Pi_N^\ast \Pi_N) \rg\kappa_x}{ \tilde{\mc{K}}^t \Pi_N^\ast \psi}|\\ 
    & \leq R_N \|\rg\kappa_x\| \|\tilde{\mc{K}}^t \Pi_N^\ast\psi\|.
\end{aligned}
\end{footnotesize}$$
    Since $\mc{K}^t$ forms a semigroup, $\mc{K}^0 =I$, there exists $c>0$ and $\alpha>0$, with $\|\mc{K}^t\|\leq ce^{\alpha t}$. When $\Pi_N^\ast\psi\in D(\mc{L})$, we have $\|\tilde{\mc{K}}^t\Pi_N^\ast\psi\| = \|\int_0^t \mc{K}^s \mc{L}\Pi_N^\ast \psi \xD{s}\| \leq (c/\alpha)(e^{\alpha t}-1)\|\mc{L}\Pi_N^\ast \psi\|$. The conclusion then follows from $\Pi_N^\ast\psi=\psi$ in $\mc{G}$. 
\end{proof}

\begin{remark}
    The existence of $R_N$ can be guaranteed by constructing an $H^s$-approximation of $\kappa(x,\cdot) = \rho(\|x-\cdot\|)$ on the compact region $\mc{X}$ (whose diameter must be finite, say $1$). In the case of polynomial approximation, this is possible with, e.g., $\kappa^N(x, \cdot)= \int_{\|\xi\|\leq 1} \kappa(x,\xi+\cdot) q_N(1-\|\xi\|^2)^N\xD{\xi}$ with $q_N$ a normalizing constant. Indeed, by change of variables it can be seen that $\kappa^N(x, \cdot)$ is a polynomial of degree $2N$ and preserves the $C^\alpha$ modulus of $\kappa(x, \cdot)$. 
\end{remark}
\begin{remark}
    In Theorem \ref{th:deterministic.error}, $R_N$ is a projection error and decreases with increasing $N$. 
    Yet, different from the invariance proximity metric \cite{haseli2023generalizing} which is evaluated over all functions on $\mc{G}^N$, $R_N$ is characterized over all canonical features of $x\in \mc{X}$. 
    As we know, $\{\rg\kappa_x: x\in \mc{X}\}$ is an $n$-dimensional manifold in $\mc{G}$. Hence our characterization is much less conservative. 
\end{remark}
\begin{remark}
    The constant $C_N$ stands for a condition number. Since $\rg H^s(\mc{X}) = \mc{G}$ is compactly embedded into $L^2(\mc{X})$, the integral of outer products $\int_\mc{X} \rg\kappa_x\times \rg\kappa_x \xD{x}$ must be an operator with decaying eigenvalues, and hence the operator norm of its inverse grows with increasing $N$. \emph{Whether} there is a tradeoff between $R_N$ and $C_N$ depends on the spectral properties of the reproducing kernel, as well as the choice of basis.\footnote{For example, one can show that if $\mc{G}^N$ is spanned exactly by the eigenfunctions of the integral operator with the kernel $\rg\kappa$ associated with the largest $N$ eigenvalues $\lambda_1 \geq \cdots \geq \lambda_N$ ($\lim_{N\to\infty} \lambda_N=0$), then $R_N\leq \lambda_N$ while $C_N\leq 1/\lambda_N$. Thus, $R_N(C_N+1)\to 1$, which implies no ``tradeoff''.} Hence, when a user-assigned dictionary is used in EDMD, it is indeed imperative to avoid ``overfitting''. 
\end{remark}

\subsection{Statistical Error of gEDMD with Finite Data}
While \eqref{eq:EDMD.concept.dt} is not tractable computationally, when the dictionary functions $\phi = (\phi_1, \cdots, \phi_N)$ are given, it is an equivalent approach to directly identify $\mc{L}_N$ from the following optimization problem:
\begin{equation}
\textstyle
    \min_{\mc{L}_N\in \mR^{N\times N}} \int_{\mc{X}} \|\dot\phi(x) - \mc{L}_N\phi(x)\|^2 \xD{x}, 
\end{equation}
whose explicit solution is given by
\begin{equation}\label{eq:EDMD.actual.ct}
\textstyle
    \mc{L}_N = \bra{\int_{\mc{X}} \dot\phi(x)\phi(x)^\top \xD{x}} \bra{\int_{\mc{X}} \phi(x)\phi(x)^\top \xD{x}}^{-1}.
\end{equation}

\par Based on the error bound in Theorem \ref{th:deterministic.error}, we can derive the error on the use of this estimation $\mc{L}_N$ to predict the evolution of the values of basis functions $\phi$, namely on the explicit lifting of $x$ on $\mr{span}\{\phi_k\}_{k=1}^n$. The following corollary can be proved easily by the fact that $\|\rg\kappa_x\| = \|x\| \rho(0)^{1/2}$ and taking the limit of $t\downarrow0$ in Theorem \ref{th:deterministic.error}. Here, we assume without loss of generality that $\mc{L}\phi_1, \cdots, \mc{L}\phi_N$ are normalized.  
\begin{corollary}\label{cor:deterministic.error}
    Under the conditions of Theorem \ref{th:deterministic.error}, we have 
    $$|\mc{L}_N\phi(x)-\dot{\phi}(x)|\leq \sqrt{N}\log\alpha \cdot R_N(C_N+1)\cdot \rho(0)^{1/2}\|x\|$$
    where $\rho$ is the radial function in the kernel (cf. Lemma \ref{lem:linear--radial}). Hence, if $\|\phi(x)\| \geq \|x\|$ (e.g., if $\phi$ contains all $e_k$), then 
    $$|\mc{L}_N\phi(x)-\dot{\phi}(x)|\leq \sqrt{N}\log\alpha \cdot R_N(C_N+1)\cdot \rho(0)^{1/2}\|\phi(x)\|$$
    holds for all $x\in \mc{X}$. 
\end{corollary} 

\par When a finite independent sample $\{x_i, \dot{x}_i\}_{i=1}^M$ (from which $\dot\phi(x_i) = \mr{D}\phi(x_i)\dot{x}_i$ can be calculated) is available, then the integrals in the estimation \eqref{eq:EDMD.actual.ct} are replaced by sample averages:
\begin{equation}\label{eq:gEDMD}
\textstyle
    \mc{L}_{N,M} = \bra{\frac{1}{M} \sum_{i=1}^M \dot\phi(x_i)\phi(x_i)^\top} \bra{\frac{1}{M} \sum_{i=1}^M \phi(x_i)\phi(x_i)^\top}^{-1} 
\end{equation}
which, as the sample size $M\rightarrow \infty$, converges to \eqref{eq:EDMD.actual.ct} in probability. It is hence possible to obtain a concentration inequality for the error between $\mc{L}_{N,M}$ and its large-data limit $\mc{L}_N$. We quote the conclusion below from Schaller et al. \cite{data_bound}. 

\begin{theorem}[Stochastic error of prediction]
\label{th:stochastic.error}
    Let $\phi_1$, $\cdots$, $\phi_N \in \rg H^s(\mc{X})\cap C^1(\mc{X})$ and sample $\{x_i\}_{i=1}^M$ be independently and uniformly distributed. 
    For any given statistical error bound $\epsilon > 0$ and tolerance $\delta\in(0,1)$, there is an amount of data $M_{N,\epsilon,\delta} \in \mathbb{N}$, such that whenever $M \geq M_{N,\epsilon,\delta}$, $\| {\cal L}_{N,M} - {\cal L}_N \| \leq \epsilon$  holds with probability $1-\delta$.\footnote{
Specifically, $ M_{N, \epsilon} \ge 
(3N^2/\tilde{\epsilon}_{r,0}^2 \delta) \max\bigl\{ \|\Sigma_1\|_{\mr{F}}^2, \|\Sigma_2\|_{\mr{F}}^2 \bigr\}$, where $\Sigma_1$ and $\Sigma_2$ are the variance matrices: 
\[\begin{aligned}
(\Sigma_1)^2_{ij}
=& \bra{|{\cal X}|^{-1} \|\phi_i\|_{L^2}^2 - \mu_i^2 } \langle\nabla \phi_j, f\rangle_{L^2}^2,\\
(\Sigma_2)^2_{ij} 
= & |{\cal X}|^{-2} \langle \phi_i^2,\phi_j^2 \rangle_{L^2}  - |{\cal X}|^{-2} \langle \phi_i,\phi_j \rangle_{L^2}^2, \enspace i,j=1,\cdots,N,
\end{aligned}\]
$|\cal X|$ is the Lebesgue measure of $\cal X$, $\mu_i = |\mc{X}|^{-1} \int_{\mc{X}} \phi_i(x) \xD{x}$, and  
$$ \tilde{\epsilon} \;=\; \min \left\{1, \|R_1 \|_{\mr{F}}^{-1} \|R_2^{-1}\|_{\mr{F}}^{-1} \right\} \cdot 
\frac{\epsilon \|R_1 \|_{\mr{F}}}{2\|R_1 \|_{\mr{F}}\|R_2^{-1}\|_{\mr{F}} + \epsilon}, $$
where matrices $R_1, R_2 \in \mathbb{R}^{N \times N}$ are given by $(R_1)_{ij}: = \langle \phi_i, {\cal L} \phi_j \rangle_{L^2}$ and $(R_2)_{ij}: = \langle \phi_i, \phi_j \rangle_{L^2}$.
}
\end{theorem}

\par Combining Corollary \ref{cor:deterministic.error} and Theorem \ref{th:stochastic.error} for the deterministic error and stochastic error respectively, we obtain the total error bound, which remains sectorial. 
\begin{corollary}[Total error of gEDMD]
\label{cor:total.error}
    Under the conditions of Theorem \ref{th:deterministic.error} and \ref{th:stochastic.error}, if the dictionary functions satisfy $\|\phi(x)\|\geq \|x\|$, then there exists a constant $c_r$, dependent on $N$, $M$, and $\delta$, such that with confidence $1-\delta$, it holds that $$\|\mc{L}_{N,M}\phi(x)- \mc{L}\phi(x)\| \leq c_r\|\phi(x)\|.$$ As such, we have ``lifted'' the system \eqref{nonlinear_system} as
    \begin{equation}
        \dot{z}(t) = \mc{L}_{N,M}z(t) + r(x(t)), 
    \end{equation}
    with remainder $\|r\| \leq c_r\|z\|$. 
\end{corollary}

\section{Koopman-based Robust Observer Design}\label{sec_KoopmanBasedObserver}
\par The purpose of theoretically establishing the remainder bound in the approximate linear lifting of the nonlinear system is to utilize it in robust \emph{state observer} synthesis. 
The fact that the remainder $r$ is a \emph{sectorial} uncertainty is crucial in the problem formulation. 
Although the formulation resembles that of \cite{SR_TAC_24}, the error analysis above has included both deterministic structural mismatch due to finite dictionary and the stochastic error due to finite data. 

\subsection{Form of the Observer}
\par The observer that we consider is specified by a Luenberger gain matrix $L\in \mR^{N\times m}$, which is to be determined:
\begin{equation}\label{eq:observer}
    \dot{\hat{z}}(t) = A\hat{z}(t) + L(y(t)-C\hat{z}(t)), 
\end{equation}
where $A = \mc{L}_{N,M}$ is identified from the gEDMD procedure in the foregoing section. Clearly, the state estimation error, $e(t) = z(t)-\hat{z}(t)$, satisfies a linear dynamics with remainder term $r(x(t))$:
\begin{equation}\label{eq:error.dynamics}
    \dot{e}(t)=(A-LC)e(t) + r(x(t)). 
\end{equation}

\par Below we make the assumption that the output mapping is an exact linear combination of the dictionary functions. This allows us to confine the error in the state dynamics, so that asymptotic decay of the state observation error is achievable (i.e., $\|e(t)\|\leq c\exp(-\alpha t)$ for some $\alpha>0$ and $c>0$); otherwise, the output mismatch must affect the observer performance, although it is not hard to capture by a uniform bound from the function approximation theory. 
Yet, this assumption is not restrictive, as the output--state relation tends to be simple (e.g., outputs are some state components), and the user may be able to incorporate the novel terms in $h$ into the dictionary. 
\begin{assumption}\label{assump2}
    $\forall i \in \{1,2,\cdots,m\}$, $h_i \in \textnormal{span} (\phi)$. That is, $\exists C\in \mathbb{R}^{m \times N}$, such that $y = C \phi(x) = Cz$. 
\end{assumption}
Then, under the conditions set in Corollary \ref{cor:total.error}, the lifted system of~\eqref{nonlinear_system} is rewritten as
\be\label{eq19} 
    \dot{z}(t) = Az(t) + r(x(t)), \quad 
    y(t) = Cz(t).
\ee
with $\|r(x(t))\|\leq c_r\|z(t)\|$. 
\begin{remark}
The constant $c_r$ in~\eqref{eq19}, which captures the total error of gEDMD in predicting the time derivative of the lifted state, is not directly computable. 
Statistically, one can resample the state space and fit to the inequality $\|r(x)\|\leq c_r\|\phi(x)\|$ (as we will do in \S\ref{sec_numerical} for numerical studies). 
In view of possible noises in the data sample, especially the data for $\dot{x}$, the constant $c_r$ can also be user-assigned. 
The choice of $c_r$ thus reflects a trade-off between the robustness of the observer and the conservativeness of the remainder estimation.
\end{remark}
\begin{remark}
    In the case where the state velocity $\dot{x}$ data is not directly accessible, and the sampling on the state space cannot occur at a high rate with a high signal-to-noise ratio, the gEDMD approach for estimating $\mc{L}$ can be replaced by a \emph{resolvent-type} estimation (see, Meng et al. \cite{meng2026resolvent}). 
    In this setting, it is assumed that a (long) trajectory is observed after initialized at each sampled state point, and a Yosida approximation of $\mc{L}$ is sought via an approximated resolvent. 
    The $c_r$ bound is then dependent on the Yosida regularization constant and the trajectory length, in addition to $N$ and $M$. 
\end{remark}

\subsection{Observer Synthesis: Asymptotically Stable Systems}
\par Now we aim to design the observer to achieve a desired exponential convergence rate $\alpha > 0$. 
If the approximate linear system \emph{$A$ is stable}, i.e., the system would converge to the origin if there were no remainder term $r$, then we have the following theorem, which establishes a robust LMI condition on two matrices $P_z$ and $P_e$ that parameterize a Lyapunov function to certify the decay of observation error. 
If the LMI is feasible, then the resulting observer $L = P_e^{-1}G$ guarantees such a desired convergence rate with confidence $1-\delta$. 

\begin{theorem}\label{theorem}
    Suppose that Assumption \ref{assump2} holds. 
    For given $\alpha>0$ and $c_r > 0$, if there exist matrices $P_z, P_e \succ 0$ in $\mR^{N \times N}$, $G \in \mR^{N \times m}$, and scalar $\lambda \ge 0$ satisfying the LMI: 
    \begin{equation}\label{eq_condi_LMI}
        \begin{small}
        \begin{bmatrix}
            (P_zA)^\ms{s} + 2\alpha P_z + \lambda c_r^2 I & 0 & P_z \\ 
            0 & (P_e A - GC)^\ms{s} + 2\alpha P_e & P_e \\
            P_z & P_e & -\lambda I
            \end{bmatrix} \preceq 0,
        \end{small}
    \end{equation}
    then the observer with $L = P_e^{-1}G$ achieves an exponential convergence rate $\alpha$ in the state observation error for system \eqref{eq19} with $\|r(x(t))\|\leq c_r\|z(t)\|$. 
    Here, we use the notation $M^\ms{s} := (M+M^\top)/2$ for a matrix $M$. 
\end{theorem}
\begin{proof}
We show that for the Lyapunov function candidate of the form $V(z, e) = z^\top P_z z + e^\top P_e e$, \eqref{eq_condi_LMI} implies that $\dot{V} \leq -2\alpha V$. Once so, we have $e(t)^\top P_e e(t)\leq V(z(t), e(t)) \leq V(z(0), e(0))e^{-2\alpha t}$ and thus $\|e(t)\|\leq \sqrt{ V(z(0),e(0))/\lambda_{\min}(P_e) } e^{-\alpha t}$. 
\par From the lifted state dynamics \eqref{eq19} and error dynamics \eqref{eq:error.dynamics}, we obtain:
$$\begin{small}
\begin{aligned}
\dot{V} &= z^\top P_z (Az+r)  + (Az+r)^\top P_z z \\
&\quad + e^\top P_e ((A-LC)e + r)  + ((A-LC)e + r)^\top P_e e \\
&=\begin{bmatrix} z \\ e \\ r\end{bmatrix}^\top \begin{bmatrix}
0&P_z&0&0\\
P_z&0&0&0\\
0&0&0&P_e\\
0&0&P_e&0
\end{bmatrix} \begin{bmatrix}
I&0&0\\
A&0&I\\
0&I&0\\
0&A-LC&I
\end{bmatrix} \begin{bmatrix} z \\ e \\ r\end{bmatrix}.
\end{aligned}
\end{small}$$ 
From the bound for $r$, we have
$$\begin{small}
\begin{bmatrix} z \\ r \end{bmatrix}^\top
\begin{bmatrix} c_r^2 I&0\\0& -I 
\end{bmatrix}
\begin{bmatrix} z \\ r \end{bmatrix} \geq 0. 
\end{small}$$
Clearly, the following condition is sufficient for $\dot{V}+2\alpha V<0$ to hold at all nonzero $(x, e)$: $\exists \lambda\ge 0$, such that for all $(z,e,r)$, 
$$\begin{small}
\begin{aligned} 
&\begin{bmatrix} z \\ e \\ r\end{bmatrix}^\top \! \begin{bmatrix}
0&P_z&0&0\\
P_z&0&0&0\\
0&0&0&P_e\\
0&0&P_e&0
\end{bmatrix} \begin{bmatrix}
I&0&0\\
A&0&I\\
0&I&0\\
0&A-LC&I
\end{bmatrix} \begin{bmatrix} z \\ e \\ r \end{bmatrix} \\
&\quad 
+ 2\alpha 
\begin{bmatrix} z \\ e \end{bmatrix}^\top \begin{bmatrix} P_z & 0 \\ 0 & P_e \end{bmatrix}
\begin{bmatrix} z \\ e \end{bmatrix} 
+ \lambda 
\begin{bmatrix} z \\ r \end{bmatrix}^\top \begin{bmatrix} c_r^2 I&0\\0& -I \end{bmatrix}
\begin{bmatrix} z \\ r \end{bmatrix} 
\le 0.
\end{aligned} 
\end{small} $$
The technique used above that introduces the auxiliary scalar $\lambda$ is well-known as the S-procedure \cite{boyd2004convex}. 
The inequality above can be rewritten as:
$$\begin{small}
\begin{aligned} 
&\begin{bmatrix} z \\ e \\ r\end{bmatrix}^\top \! \begin{bmatrix}
0&P_z&0&0\\
P_z&0&0&0\\
0&0&0&P_e\\
0&0&P_e&0
\end{bmatrix} \begin{bmatrix}
I&0&0\\
A&0&I\\
0&I&0\\
0&A-LC&I
\end{bmatrix} \begin{bmatrix} z \\ e \\ r \end{bmatrix} \\
&\quad + 2\alpha 
\begin{bmatrix} z \\ e \\ r \end{bmatrix}^\top 
\begin{bmatrix} I & 0 & 0 \\ 0 & I & 0 \end{bmatrix}
\begin{bmatrix} P_z & 0 \\ 0 & P_e \end{bmatrix}
\begin{bmatrix} I & 0 \\ 0 & I \\ 0 & 0 \end{bmatrix}
\begin{bmatrix} z \\ e \\ r \end{bmatrix}
\\
&\quad + \lambda 
\begin{bmatrix} z \\ e \\ r \end{bmatrix}^\top 
\begin{bmatrix} I & 0 & 0 \\ 0 & 0 & I \end{bmatrix}\begin{bmatrix} c_r^2 I&0\\0& -I \end{bmatrix}
\begin{bmatrix} I & 0 \\ 0 & 0 \\ 0 & I \end{bmatrix}
\begin{bmatrix} z \\ e \\ r \end{bmatrix} \le 0.
\end{aligned}
\end{small} $$
Thus, a sufficient condition for $\dot{V}<0$ at all $(z, e) \neq 0$ is 
$$\begin{small}
\begin{aligned}
\begin{bmatrix}
  (P_zA)^\ms{s} + 2\alpha P_z+ \lambda c_r^2 I & 0 & P_z\\
  0 & (P_e(A-LC))^\ms{s} + 2\alpha P_e &  P_e\\
  P_z & P_e & -\lambda I
\end{bmatrix} \preceq 0,
\end{aligned}
\end{small}
$$
which is equivalent to \eqref{eq_condi_LMI} with $G: = P_e L$. 
\end{proof}

\subsection{Observer Synthesis: Systems without Asymptotic Stability}
If the system is not asymptotically stable, i.e., if $z\rightarrow 0$ cannot hold with the identified $A$ even without the remainder $r$, then it is impossible to have a Lyapunov function candidate that is a quadratic form of $(z,e)$. In such cases, as long as $x(t)$ by our standing assumption is in the bounded region $\mc{X}$, then $z(t)$ remains bounded and hence $r$ is bounded too. 
It is thus of interest to find an observer with a \emph{minimum guaranteed $L^2$-gain from $r$ to the observer error $e$.} 
In particular, we seek to find a $P_e\succeq 0$ that defines a storage function $V(e)=e^\top P_ee$, that is dissipative:
\begin{equation}\label{eq:dissipation}
    \dot{V} \leq -\|e\|^2 + \gamma \|r\|^2.
\end{equation} 
As such, we obtain $\int_0^T \|e(t)\|^2 \xD{t}\leq \gamma \int_0^T\|r(t)\|^2 \xD{t} + e(0)^\top P_ee(0)$, which implies an $L^2$-gain not exceeding $\gamma^{1/2}$. 

\begin{theorem}
    Suppose that Assumption \ref{assump2} holds. 
    For given $\gamma\geq 0$ and $c_r > 0$ if there exist matrices $P_e \succeq 0$ in $\mR^{N \times N}$ and $G \in \mR^{N \times m}$ satisfying the LMI: 
    \begin{equation}\label{eq_condi_LMI2}
        \begin{small}
        \begin{bmatrix}
            (P_e A - GC)^\ms{s} + I & P_e \\
            P_e & -\gamma I
            \end{bmatrix} \preceq 0.
        \end{small}
    \end{equation}
    then \eqref{eq:dissipation} holds for $V(e)=e^\top P_e e$, and hence the $L^2$-gain from $r$ to $e$ is upper bounded by $\gamma^{1/2}$. 
\end{theorem} 
\begin{proof}
    Clearly, $\dot{V}+\|e\|^2 - \gamma\|r\|^2 = e^\top P_e((A-LC)e+r) + ((A-LC)e+r)^\top P_e e + \|e\|^2 - \gamma\|r\|^2$ is a quadratic form of $(e, r)$. The matrix that defines this quadratic form is the one in \eqref{eq_condi_LMI2}. 
\end{proof} 

Therefore, the $L^2$-optimal observer can be obtained by searching for $P_e, G, \gamma$ subject to \eqref{eq_condi_LMI2} with the objective of minimizing $\gamma$. We summarize the proposed algorithm for observer design in Algorithm~\ref{alg1}.
\begin{algorithm}[t]
\caption{gEDMD-based robust observer synthesis}\label{alg1}
\begin{algorithmic}[1]
\Statex \textbf{Input:} Data $\{x_i,\ \dot x_i\}_{i=1}^M$; Dictionary $\phi = (\phi_1, \cdots, \phi_N)$; Sectorial error bound $c_r>0$.
\State Calculate $\phi(x_i)$ and $\dot{\phi}(x_i) = \mr{D}\phi(x_i)\dot{x}_i$ for $1\le i\le M$;
\State Obtain matrix $A=\mc{L}_{N,M}$ via gEDMD \eqref{eq:EDMD.actual.ct};  
\If{$A$ is stable}
  \State Solve the LMI \eqref{eq_condi_LMI} to obtain $P_e$ and $G$. 
  \Else 
  \State Minimize $\gamma$ subject to  \eqref{eq_condi_LMI2} to obtain $P_e$ and $G$. 
\EndIf 
\Statex \textbf{Output:} Observer gain $L = P_e^{-1} G$.
\end{algorithmic}
\end{algorithm}

\section{Numerical Examples}\label{sec_numerical}
We illustrate the effectiveness of the proposed observer design through numerical simulations. All experiments are implemented in Python using CVXPY \cite{diamond2016cvxpy} for optimization. The codes are available at the authors' \href{https://github.com/XiuzhenYe/EDMD-Based-Robust-Observer/}{GitHub repository}.% The simulations demonstrate the observer performance and provide insights into robustness and practicality.

\subsection{A Stable System with Invariant Koopman Lifting}
Consider an asymptotically stable nonlinear system \cite{BS_arXiv_21}:
$$ \dot{x}_1 = -2x_1 , \ \ \ \dot{x}_2  = -x_2  + x_1 ^2.$$
To obtain an EDMD model, we define the dictionary by
$\phi(x) = \left[x_1 \ \ \ x_2 \ \ \ x_2 +x_1^2/3 \right]^\top$.
This choice of dictionary in fact yields an exact $N$-dimensional system ($N=3$):
$$ \dot{z}(t) = Az(t), \enspace A = \begin{bmatrix}
-2 & 0 & 0\\ 0 & -4 & 3 \\ 0 & 0 & -1
\end{bmatrix} .$$
Due to such exactness, a small sample ($M = 100$) uniformly drawn from ${\cal X} = [-1, 1]^2$ suffices to yield an $\mc{L}_{N,M}$ equal to $A$ (with only numerical precision errors) through gEDMD \eqref{eq:EDMD.actual.ct}. We can thus choose a very small $c_r$, e.g., $0.001$. 

\begin{figure} 
    \centering
    \begin{subfigure}[t]{0.48\linewidth}
        \centering
        \includegraphics[width=\linewidth]{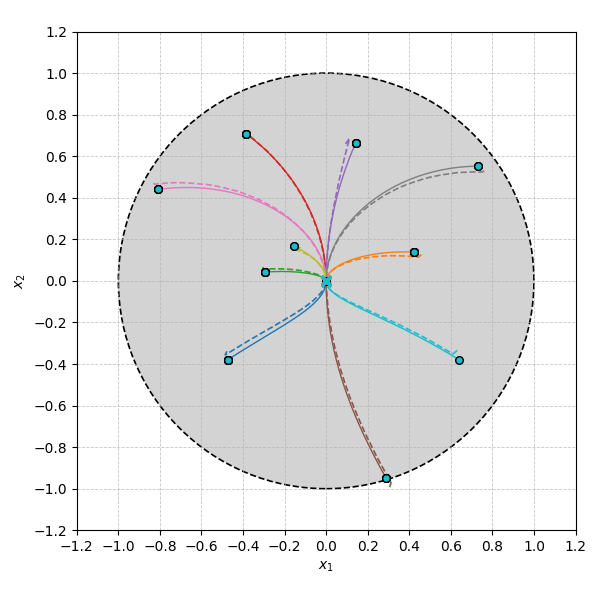}
        \caption{True state and observed state trajectories when ($\alpha=0.1$).}
        \label{fig1}
    \end{subfigure}
    \vspace{4mm} 
    \begin{subfigure}[t]{0.48\linewidth}
        \centering
        \includegraphics[width=\linewidth]{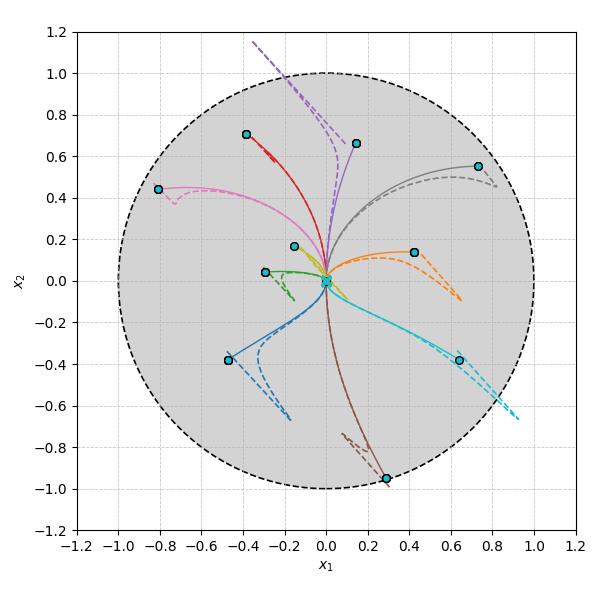}
        \caption{True state and observed state trajectories ($\alpha=0.99$).}
        \label{fig2}
    \end{subfigure}
    \centering
    \begin{subfigure}[t]{0.48\linewidth}
        \centering
        \includegraphics[width=\linewidth]{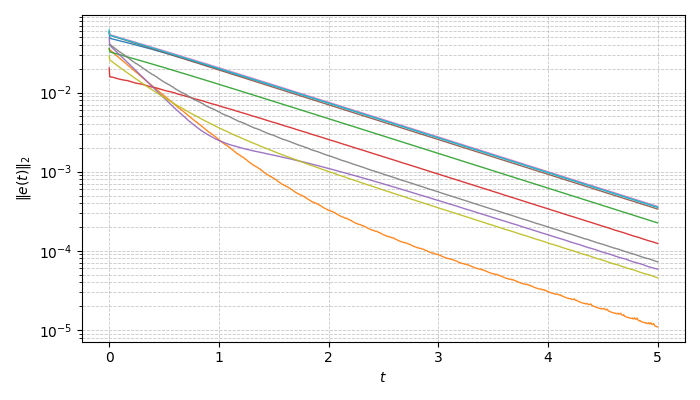}
        \caption{Error decay ($\alpha=0.1$).}
        \label{fig3}
    \end{subfigure}
    \vspace{4mm} 
    \begin{subfigure}[t]{0.48\linewidth}
        \centering
        \includegraphics[width=\linewidth]{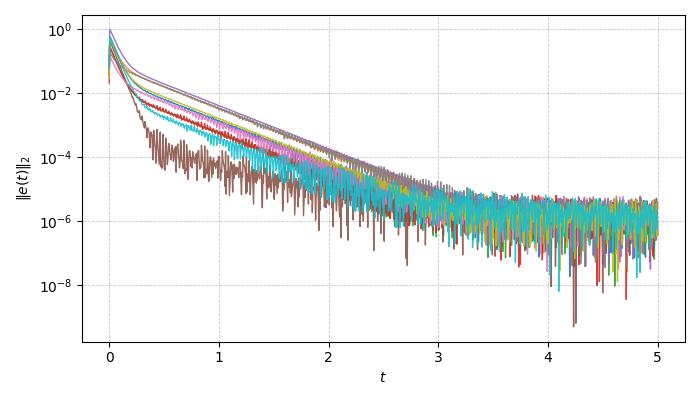}
        \caption{Error decay ($\alpha=0.99$).}
        \label{fig4}
    \end{subfigure}
    \caption{Observer performance for the system with an invariant Koopman lifting.}
    \label{fig_case1}
\end{figure}

We suppose that the output matrix $C=[1 \ 1 \ 0]$, i.e., $y = x_1+x_2$. 
Fig.~\ref{fig1} and Fig.~\ref{fig2} illustrate the trajectories of true state and observer estimates from $10$ random initial points for $\alpha = 0.1$ and $\alpha = 0.99$. (The problem should be feasible when $\alpha\leq 1$.) 
Fig.~\ref{fig3} and Fig.~\ref{fig4} depict the corresponding decay of the error norm. As expected, with larger $\alpha$, the observer converges more aggressively to the steady states at the expense of larger error (``peaking effect") in the incipient stages. 
These results confirm the feasibility of the proposed observer synthesis algorithm. 

\subsection{A Stable System without Invariant Koopman Lifting}
For a chemical process with two tank reactors in series:
$$\begin{small}
    \begin{aligned}
        \xDD{C_1}{t} =& \frac{F_1}{V_1} (C_{10} - C_1) - k_1C_1^2, \\
        \xDD{C_2}{t} =& \frac{F_2}{V_2} (C_{20} - C_2) + \frac{F_1}{V_2} (C_1 - C_2) - k_2C_2^2.
    \end{aligned}
\end{small}$$ 
with model parameters listed in Tab. \ref{table}, we define the dictionary by $\phi(x) = \left[x_1 \ x_2 \ x_1^2 \ x_2^2 \ x_1x_2 \right]^\top$, where $x_1$ and $x_2$ refer to the deviations from the steady-state values of $C_1$ and $C_2$ (denoted as $C_1^\text{ss}$ and $C_2^\text{ss}$), respectively. 
In this example, the dictionary is not invariant, i.e., there exist state-dependent functions in the span of these dictionary functions but escape therefrom after applying the Koopman operator.  
\begin{table}[!t]
\centering
\begin{tabular}{ll}
\toprule 
$F_1 = F_2 = \SI{5.0}{\meter^3\per\hour}$  & $V_{1} = V_{2} = \SI{1.0}{\meter^3}$ \\ 
$C_{10} = \SI{4.0}{\kilo\mole\per\meter^3}$ & $C_{20} = \SI{4.5}{\kilo\mole\per\meter^3}$ \\
$C_1^\text{ss} = \SI{2.0}{\kilo\mole\per\meter^3}$ & $C_2^\text{ss} = \SI{3.0}{\kilo\mole\per\meter^3}$ \\
$k_1 = \SI{2.5}{\meter^3\per\kilo\mole\per\hour}$ & $k_2 = \SI{0.2778}{\meter^3\per\kilo\mole\per\hour}$ \\
\bottomrule
\end{tabular}
\caption{Parameters of the two-CSTR-in-series}\label{table}
\end{table}

\par
\begin{figure} 
    \centering
    \begin{subfigure}[t]{0.46\linewidth}
        \centering
        \includegraphics[width=\linewidth]{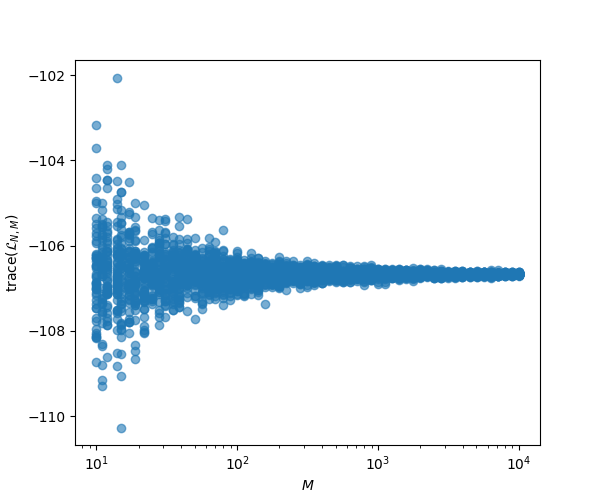}
        \caption{Effect of sample size on gEDMD.}
        \label{fig:ex2_M}
    \end{subfigure}
    \begin{subfigure}[t]{0.46\linewidth}
        \centering
        \includegraphics[width=\linewidth]{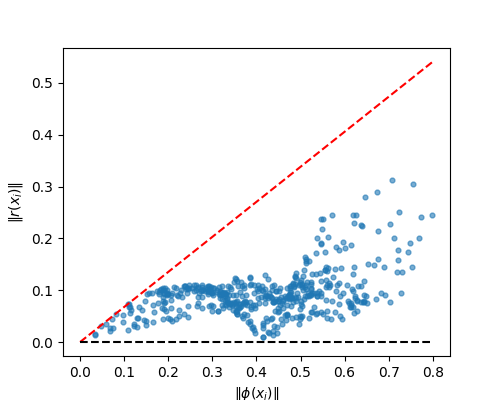}
        \caption{Scatter plot of $\|r(x)\|$ and $\|\phi(x)\|$ at the sample points.}
        \label{fig:ex2_cr}
    \end{subfigure}
    \caption{Tuning of sample size $M$ and error bound $c_r$.}
\end{figure}

 To determine an appropriate sample size, we let $M$ vary from $10^1$ to $10^4$ and identify $\mc{L}_{N,M}$ in $50$ experiments under each $M$. The sample comes from ${\cal X}= [-0.5, 0.5]^2$. 
The variation of the resulting $\mr{trace}(\mc{L}_{N,M})$ is shown in Fig. \ref{fig:ex2_M}, based on which we choose $M=500$ for a satisfactorily small stochastic error. 
As an estimation of the error of gEDMD, we use $c_r = \max\{ \|r(x_i)\|/\|\phi(x_i)\| \}_{i=1}^M$, whose value is approximately $0.67$. This sectorial error bound is confirmed by the plot in Fig. \ref{fig:ex2_cr} (indicated by the red dashed curve). 

\begin{figure} 
    \centering
    \begin{subfigure}[t]{0.46\linewidth}
        \centering
        \includegraphics[width=\linewidth]{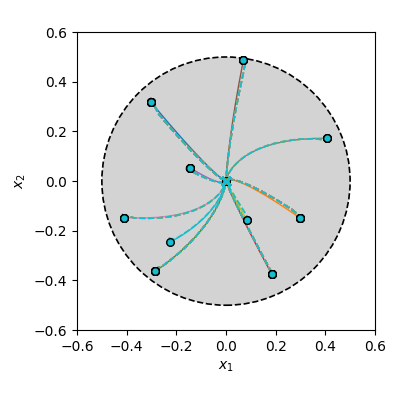}
        \caption{True state and observed state trajectories ($\alpha=5$).}
        \label{fig5}
    \end{subfigure}
    \vspace{4mm} 
    \begin{subfigure}[t]{0.46\linewidth}
        \centering
        \includegraphics[width=\linewidth]{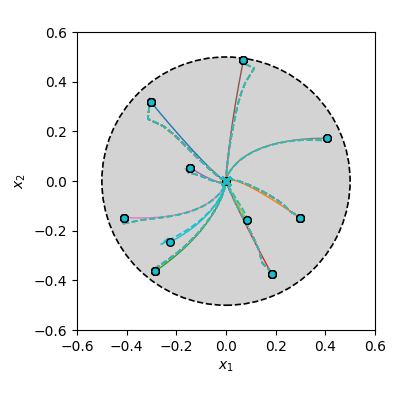}
        \caption{True state and observed state trajectories ($\alpha=10$).}
        \label{fig6}
    \end{subfigure}
    \centering
    \begin{subfigure}[t]{0.48\linewidth}
        \centering
        \includegraphics[width=\linewidth]{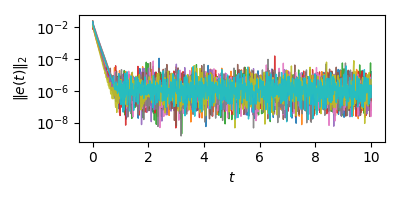}
        \caption{Error decay ($\alpha=5$).}
        \label{fig7}
    \end{subfigure}
    \vspace{4mm} 
    \begin{subfigure}[t]{0.48\linewidth}
        \centering
        \includegraphics[width=\linewidth]{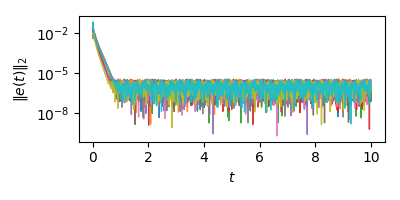}
        \caption{Error decay ($\alpha=10$).}
        \label{fig8}
    \end{subfigure}
    \caption{Observer performance for the system without an invariant Koopman lifting.}
    \label{fig_case2}
\end{figure}

We define the output matrix as $C = [0 \ 1 \ 1 \ 0 \ 0]$, i.e., $y = x_2+ x_1^2$. We simulate the CSTRs from $10$ random initial states and evaluate the observer from Algorithm~\ref{alg1}. 
Fig.~\ref{fig5} and Fig.~\ref{fig6} show the trajectories of the true state together with state estimates for $\alpha = 5$ (with a mild observer) and $\alpha = 10$ (with an overtuned observer), respectively. The error norm decays are shown in Fig.~\ref{fig7} and Fig.~\ref{fig8}. 
Even when the dictionary in gEDMD is not invariant, due to the sectorial error bound, the synthesized observer still guarantees that the estimation errors decay to negligibly low levels (below $10^{-4}$). 

\vspace{-1mm}
\subsection{A Limit Cycle System}
In the last experiment, we consider a van der Pol oscillator:
$$ \dot{x}_1 = x_2, \enspace \dot{x}_2 = 0.5(1-9x_1^2)x_2 - x_1$$
on $\mc{X} = [-1, 1]\times [-1, 1]$. The equilibrium point at the origin is unstable and the system exhibits a limit cycle. Nevertheless, the theoretical setting of this paper is satisfied, and hence gEDMD is anticipated to result in a sectorial error bound. In the unstable case, we adopt the minimal $L^2$-synthesis formulation \eqref{eq_condi_LMI2}. 
The complexity in applying our proposed gEDMD-based observer synthesis is that since the limit cycle is nontrivial, the dictionary functions, if chosen to be monomials, must span polynomials up to a sufficiently high degree. We denote by $d_{\max}$ as the maximum degree of the monomial dictionary functions. 

\par
\begin{figure} 
    \centering
    \begin{subfigure}[t]{0.44\linewidth}
        \centering
        \includegraphics[width=\linewidth]{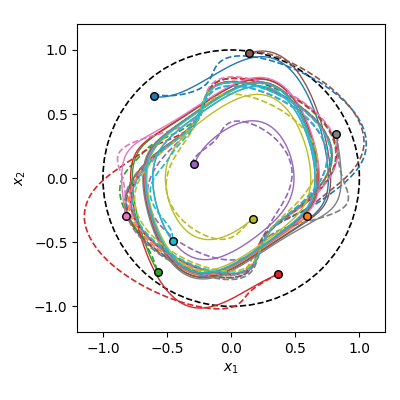}
        \caption{True state and observed state trajectories ($d_{\max}=4$).}
        \label{fig9}
    \end{subfigure}
    \vspace{4mm} 
    \begin{subfigure}[t]{0.44\linewidth}
        \centering
        \includegraphics[width=\linewidth]{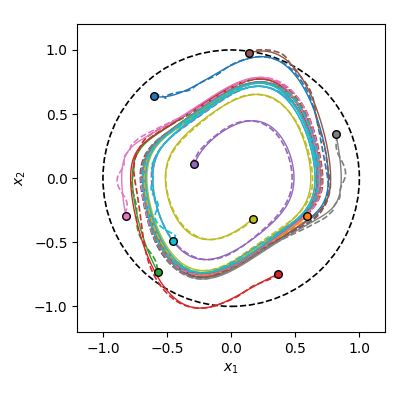}
        \caption{True state and observed state trajectories ($d_{\max}=8$).}
        \label{fig10}
    \end{subfigure}
    \centering
    \begin{subfigure}[t]{0.48\linewidth}
        \centering
        \includegraphics[width=\linewidth]{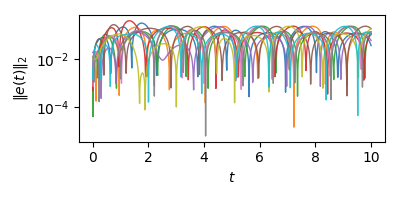}
        \caption{Error norms ($d_{\max} = 4$).}
        \label{fig11}
    \end{subfigure}
    \vspace{4mm} 
    \begin{subfigure}[t]{0.48\linewidth}
        \centering
        \includegraphics[width=\linewidth]{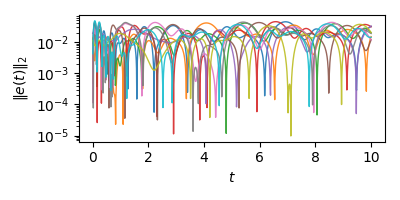}
        \caption{Error norms ($d_{\max} = 8$).}
        \label{fig12}
    \end{subfigure}
    \caption{Observer performance for the limit cycle system.}
    \label{fig_case3}
\end{figure}

 Intuitively, a non-circle limit cycle should be captured by a polynomial of degree at least $4$. Following the same tuning procedure as in the previous study, we choose $M=2000$ as a large enough sample size that settles down the stochastic error. 
With these $14$ basis functions in the dictionary, gEDMD results in a $c_r$ estimate of $2.62$. The synthesized observer has a $L^2$-gain from $r$ to $e$ bounded by $1.81$. 
The performance of the observer is visualized in Fig. \ref{fig9}, where the observer's estimated state trajectories deviate from the true trajectories, and the estimated states are attracted to a limit cycle that has a discernible distortion of the actual limit cycle. 
The RMSE (root mean square error) is $0.123$ over the $10$ simulations (see Fig. \ref{fig11}). 

\par We thus increase $d_{\max}$ to higher values, and observe that satisfactory performance is obtained at $d_{\max}=8$. 
In this case, the $c_r$ is estimated to be $1.00$ and the $L^2$-gain is minimized to $2.40$. 
Fig. \ref{fig10} shows the resulting trajectories, where the estimated states track the true states well. The RMSE is calculated as $0.022$ (see Fig. \ref{fig12}).

\vspace{-1mm} 
\section{Conclusions}\label{sec_conclusion}
In this paper, the problem of designing a state observer based on a Koopman-like surrogate model identified from gEDMD is addressed. Through theoretical analysis on a linear--radial kernel-based RKHS, the gEDMD is found to result in a lifted approximate linear model with sectorially bounded mismatch. 
As such, the observer synthesis boils down to finding a Luenberger gain matrix that achieves a prescribed convergence rate or minimizes $L^2$-gain from the mismatch, formulated as an LMI condition. 
Numerical experiments on three systems with/without asymptotic stability show that the synthesized observer has expected performance. 
Therefore, this work provides an ``elementary method'' (without explicit use of operator calculus) for data-driven observer synthesis. 

\par Now that the Koopman model can be essentially free of mismatch on an infinite-dimensional RKHS, we are naturally tempted to consider the problem of establishing a method of optimal observer design without the error caused by the preassigned dictionary. It is expected to appear as a kernel-based method. 
Of course, a deeper issue that exists with the data-driven methods for state observation is the possibility of totally lacking state data even for identifying a Koopman model or training the observer. To this end, one must first address the problem of identifying a Koopman model of nonlinear systems with input and output data only. 

\vspace{-1mm}
\bibliographystyle{ieeetr}
\bibliography{mybib}

\end{document}